\documentclass[iop,apj]{emulateapj}

\newcommand       \mum          {\,{\rm \mu m}}
\newcommand{\lapprox}{$_<\atop{^\sim}$}

\begin{document}
   \title{Comparison of the dust and gas radial structure \\
  in the transition disk [PZ99] J160421.7-213028} 
\author{Ke Zhang \altaffilmark{1},  Andrea Isella \altaffilmark{1} ,  John M. Carpenter \altaffilmark{1} , Geoffrey A. Blake \altaffilmark{2}
}

\altaffiltext{1}{Division of Physics, Mathematics \& Astronomy, MC 249-17, California Institute of Technology, Pasadena, CA 91125, USA}
\altaffiltext{2}{Division of Geological \& Planetary Sciences, MC 150-21, California Institute of Technology, Pasadena, CA 91125, USA}
              \email{Corresponding Author: kzhang@caltech.edu}
 
  \begin{abstract}
  
  We present ALMA observations of the 880\,$\mum$ continuum and CO $J $= 3-2 line emission from the transition disk around
[PZ99] J160421.7-213028, a solar mass star in the Upper Scorpius OB association.  Analysis of the continuum data indicates that 80\% of the dust mass is concentrated in an annulus extending between 79 and 114\,AU in radius. Dust is robustly detected inside the annulus, at a mass surface density 100 times lower than that at 80\,AU. The CO emission in the inner disk also shows a significantly decreased mass surface density, but we infer a cavity radius of only 31\,AU for the gas.  The large separation of the dust and gas cavity edges, as well as the high radial concentration of millimeter-sized dust grains, is qualitatively consistent with the predictions of pressure trap models that include hydrodynamical disk-planet interactions and dust coagulation/fragmentation processes. 

  \end{abstract}

   \keywords{
                Stars: pre-main sequence; planetary systems: protoplanetary discs; submillimeter: stars
               }

   \maketitle


\newpage

\section{Introduction}
\label{sec:intro}

Planets form in the disks orbiting young stars, but the paths by which the primordial gas and  dust accumulate into planetary bodies remain unclear.  The capabilities of new optical, infrared, and (sub)-millimeter telescopes can place constraints on the planet formation process by mapping the gas and dust emission of planetary systems in the act of formation. Transition disks, defined by their significantly reduced  infrared emission at wavelengths $<$8\,$\mu$m compared to the median disk emission \citep{Strom89, Wolk96}, are of particular interest.  The relative lack of infrared 
 emission implies the absence of warm dust  in the innermost disk. This dust depletion might be a  manifestation of the early stages of planet formation as a result of the dynamic interactions between the disk and forming giant planets \citep{Lin79, Artymowicz94}, but other mechanisms can suppress near-infrared dust signatures without the presence of planets, such as a decrease in the dust opacity due to grain growth  \citep[e.g.,][]{Strom89,Dullemond05} or disk photoevaporation driven by  stellar radiation \citep[e.g.,][]{Clarke01,Alexander06a}. 

The three mechanisms outlined above can reproduce the infrared Spectral Energy Distributions (SEDs) characteristic of transition disks \citep{Alexander06b, Birnstiel12, Alexander14}, but make different predictions about the relative spatial distributions of the gas and dust. In the planet-disk interaction scenario, the gaseous disk is expected to be truncated near the planet orbital radius. Depending on the viscous properties of the disk, dust grains larger than a few millimeters in size are expected to decouple from the gas and concentrate in a narrow ring with a radius larger than the gas truncation radius \citep{Zhu12, Pinilla12}.  In the grain growth scenario, the drop in infrared opacity is due to the coagulation of sub-micron grains into larger bodies, but the gaseous disk should extend inward down to a few stellar radii. Finally, disk photoevaporation preferentially removes the gaseous disk and sub-micron sized grains entrained in the gas flow.  The inner edge of the gas disk moves steadily outwards and the strong inward pressure gradient at the evaporation radius drives all dust grains smaller than a few millimeter size outwards. As a result, the grains are swept up by the moving edge of the gas disk.  In this case, the truncation radius of the gaseous and dusty disk components are expected to be similar, and some large grains ($\ge$ cm-sized) might be able to survive inside the gas-free regions if not removed by other means \citep{Alexander07}.   
 
High spatial resolution images of the gas and dust emission are thus key to investigating the nature of transition disks. Sub-millimeter observations are particularly valuable since the dust continuum traces the distribution of millimeter-sized dust grains while molecular line emission traces that of the gas.   Sub-arcsecond continuum images obtained for the brightest transition disks have revealed  large cavities and, in some cases, azimuthal asymmetries in the distribution of millimeter-sized grains \citep{Pitu07, Brown08, Hughes09, Isella10, Andrews11, Isella12, Casassus13, vanderMarel13, Isella13, Perez14}.  The CO emission has been imaged at high angular resolution in only a few cases and reveals that molecular gas is present in the dust cavity \citep{Pitu07, Isella10, Casassus13, Perez14, Bruderer14}. However, the limited sensitivity and angular resolution of existing observations, and the large optical depth of low-$J$ CO transitions, have so far hampered a detailed comparison between the gas and dust spatial distributions.

In this paper, we present  880\,$\mu$m ALMA observations of the [PZ99] J160421.7-213028 (henceforth J1604-2130) transition disk that resolve the dust continuum and molecular gas emission on subarcsecond angular scales. J1604-2130 is a member of the Upper 
Scorpius OB association at a distance of 145\,pc \citep{deZeeuw99}, with a spectral type of K2, a stellar mass of 1.0\, M$_\odot$ and an estimated age between 5 and 11\,Myr \citep{Preibisch02, Pecaut12}. 
The disk is the most massive known around K-M type stars in the Upper Scorpius OB association  \citep{Mathews12b, Carpenter14} and has one of the largest cavities reported in continuum emission -- R$\sim$72\,AU, as revealed by 880\,$\mu$m SMA observations \citep{Mathews12}. The SMA observations also suggest that the CO $J$=3-2 distribution that may extend closer to the star than the millimeter-emitting dust. Our ALMA observations achieve substantially higher signal-to-noise ratios compared to the SMA observations to place stringent constraints on the spatial distributions of gas and dust in J1604-2130. 

This paper is organized as follows.  The observations and data calibration are presented in Section~\ref{sec:obs}. Section~\ref{sec:results} discusses the morphology of the dust and gas emission. The ALMA data are compared to the predictions of theoretical models for the disk emission in Section~\ref{sec:model}, and the results of the model fitting are presented at the end of the same section. In Section~\ref{sec:dis}, we discuss the nature of the J1604-2130 transition disks and a summary of the main results follows in Section~\ref{sec:summary}.

\section{Observations}
\label{sec:obs}

The observations of J1604-2130 were carried out on 28 August 2012 
as part of  the ALMA cycle 0 project 2011.0.00526.S  \citep{Carpenter14}
with 28 12-meter antennas. The total integration time on J1604-2130 was 315 seconds.  
Baselines ranged from 21 to 402\,m (23-457\,k$\lambda$).  
The spectral setup consists 
of four windows centered at 345.77, 347.65, 335.65, and 333.77\,GHz, respectively,
each providing a total bandwidth of 1.875\,GHz in 488 kHz (0.42\,km s$^{-1}$) channels. The data are Hanning smoothed, 
so the spectral resolution is twice the individual channel width. 

Reduction and calibration of the data were performed using the Common Astronomy Software 
Application (CASA) version 4.1 \citep{McMullin07}. Quasar J1625-2527 was observed 
before and after the on-source integrations for phase calibration. Short-term phase variations 
were corrected using the on-board water vapor radiometers, and the visibilities were 
self-calibrated on the continuum emission of the science source to reduce atmospheric decoherence.  
The bandpass was calibrated with the radio sources J1751+0939 and J1924-292, and the visibility amplitude scale was calibrated from observations 
of Titan with an estimated uncertainty of $\sim$10\%. For  each $uv$ point, we fitted the visibilities of each spectral window with a 
power-law ($F_\nu\propto\nu^\alpha$). We then subtracted the best-fit result from the data,  and measured the variance ($\sigma^2$) in the residual; $1/\sigma^2$ is used as the weight of each $uv$ point. 

Four lines are detected: CO\,$J$=\,3-2 (345.795\,GHz), H$^{13}$CO$^+$ $J$=\,4-3 (346.998\,GHz), 
H$^{13}$CN $J$=\,4-3 (345.339\,GHz),  and HN$^{13}$C $J$=\,4-3 (348.340\,GHz).
Continuum emission from J1604-2130 was extracted from the total 
7.5\,GHz coverage by combining the line-free channels from all four windows after 
flagging the edge channels. The spectral lines were first continuum-subtracted in 
the $uv$ domain, and the composite visibilities then Fourier inverted, deconvolved 
with the CLEAN algorithm using natural weighting, and restored 
with a synthesized beam of size 0\arcsec.73$\times$0\arcsec.46 
(106$\times$67\,AU at 145\,pc). The beam major axis position angle is 105$^{\circ}$ east of north. 
Here we focus on an analysis of the dust continuum and CO data because the optically thick CO emission better traces 
the full radial extent of the gaseous disk.

\section{Results}
\label{sec:results}

\begin{figure*}[htbp]
\begin{center}
\vspace{-0.8cm}
\includegraphics[width=\textwidth]{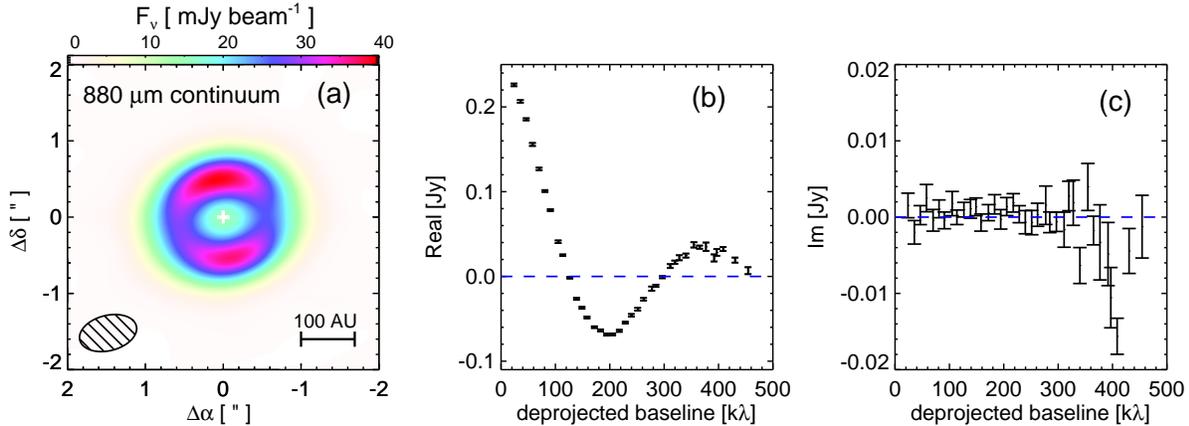}
\caption{ (a) 880\,$\mum$ continuum map of J1604-2130.  The synthesized beam, shown at lower left, has a FWHM of 0.\arcsec73$\times$0.\arcsec46, while the wedge marks the conversion from color to surface brightness. The rms noise level is 0.18\,mJy beam$^{-1}$. (b) Azimuthally averaged real part of the correlated flux  as a function of deprojected baseline length, calculated by assuming a disk inclination of 6\arcdeg\ and a position angle of -13\arcdeg \citep{Mathews12}. (c)  Azimuthally averaged imaginary 
part of the correlated flux  as a function of deprojected baseline length. }
\label{fig:cont}
\end{center}
\end{figure*}

Figure~\ref{fig:cont} presents the 880\,$\mum$ synthesized continuum image of J1604-2130 along with the real and imaginary visibilities. The continuum image shows a resolved ``ring" morphology, characterized by a radius of about 0.\arcsec6, or 90~AU at 145\,pc. The continuum flux density integrated over a central 4\arcsec~square box is 226$\pm$1\,mJy (with 10\% absolute flux uncertainty).  \cite{Mathews12} reported a lower continuum flux of 165$\pm$6 mJy (20\% absolute flux uncertainty) from SMA observations by integrating all flux above a 2$\sigma$ contour.  Using an integration region similar to that in \citet{Mathews12}, we obtained a total flux of 202\,mJy, consistent with the SMA result within the absolute flux uncertainties.   The high signal-to-noise (S/N) ALMA image allows the inclusion of weaker emission from the outer disk, and is therefore a more accurate measure of  the integrated continuum flux density. The synthesized continuum image has an rms noise level of 0.18\,mJy beam$^{-1}$, yielding a signal-to-noise ratio of $>$200.

The radial profile of the real visibilities shows two nulls at 120\,$k\lambda$ and 300\,$k\lambda$, respectively. The amplitudes of the first and second lobe are about 30\% and  13\% of the intensity on the shortest baselines, indicating a sharp transition in the radial surface brightness profile. Similar structure is seen in the SMA observations of J1604-2130 but with a lower amplitude, suggesting a smoother radial profile. The difference in morphology between the ALMA and SMA observations is possibly due to the fact that the higher sensitivity of ALMA observations allowed self-calibration to be applied to the data, reducing the de-correlation caused by atmospheric turbulence. 

The radially averaged imaginary visibilities show values consistent with zero up to about 400~$k\lambda$, implying that the continuum emission is symmetric on angular scales as small as 0.\arcsec5. Note that the apparent double-lobed morphology in the continuum map  is due to the elliptical shape of the beam, and not to an intrinsic asymmetry in the dust emission.  A slight deviation from azimuthal symmetry is observed on the longest baselines, where the imaginary visibilities reach values near 10 mJy, and
in the continuum image, where  the northern peak is $\sim$10\% brighter than its counterpart to the south. 

\begin{figure*}[htbp]
\begin{center}
\vspace{-0.6cm}
\includegraphics[width=7.2in]{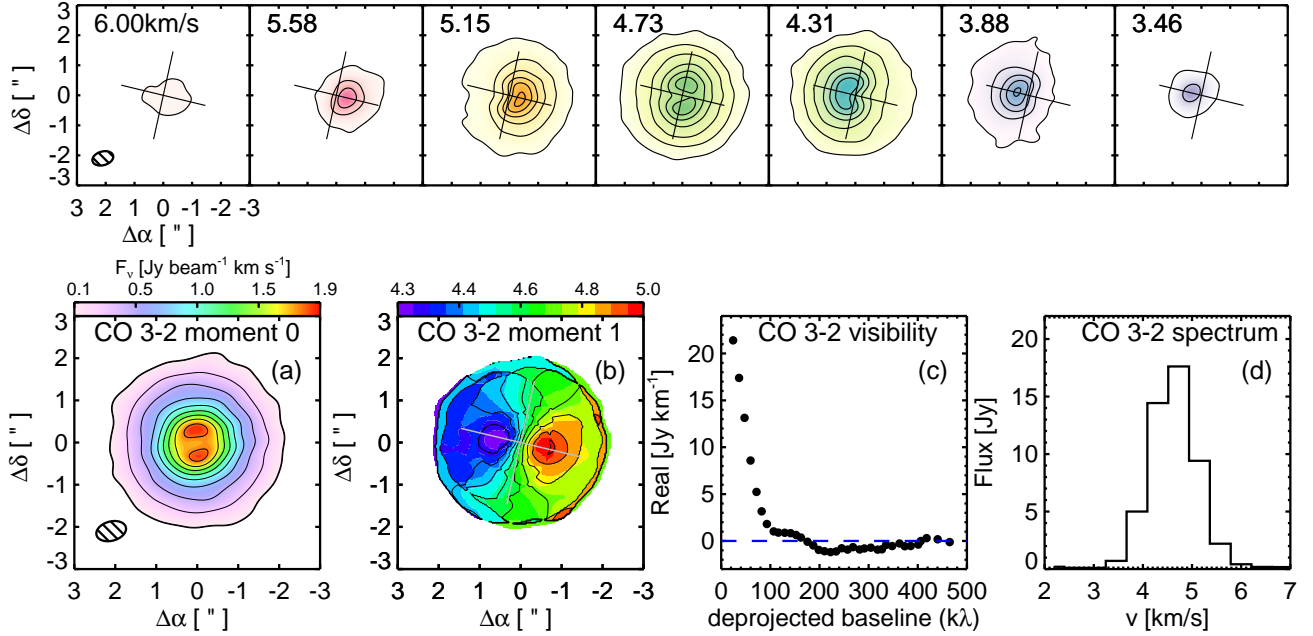}

\caption{Top row:  channel maps of the CO $J$ = 3-2 emission from J1604-2130.  
The contours begin at 3$\sigma$ (45\,mJy beam$^{-1}$) in each 0.42\,km s$^{-1}$ wide channel 
and are spaced by 10$\sigma$. The cross indicates the direction of the disk major and minor axes.
Bottom row: The moment 0 and moment 1 maps of CO $J$ = 3-2 emission are shown in (a) and (b). 
Contours on the moment 0 map start at the 3$\sigma$ level and increase by 10$\sigma$;  those
on the moment 1 map are spaced by 0.08\,km s$^{-1}$. 
(c) Azimuthally averaged CO $J$ = 3-2 emission visibilities as a function of deprojected baseline length, 
integrated over the seven channels plotted in the upper row. (d) CO $J$ = 3-2 spectrum of J1604-2130 integrated over
a central 6\arcsec$\times$6\arcsec box. }
\label{fig:co}
\end{center}
\end{figure*}

An overview of the CO $J$ = 3-2 data is provided in Figure~\ref{fig:co}. The line channel maps in the first row show clear signatures of rotation, with emission firmly detected ($>$5$\,\sigma$) over a narrow velocity range between 3.46 and 6 km s$^{-1}$ due to the nearly face on geometry. We measure an integrated intensity of 21.4$\pm$0.2\,Jy km s$^{-1}$ over a central 6\arcsec~square box, consistent with CO 3-2 line flux from single dish observations of 21.7$\pm$0.8\,Jy km s$^{-1}$ \citep{Mathews13} but 4.1$\times$ larger than that reported by the SMA \citep{Mathews12}, for which the line flux was calculated by summing over only those regions with a signal-to-noise ratio$\geqslant$2. The higher S/N of the ALMA image allowed CO $J$=3-2 emission to be detected over an extended region of the disk which contains a significant fraction of the total flux.

In the second row of Figure.~\ref{fig:co} we display the CO $J$ = 3-2 zeroth\,(a) and first\,(b) moment maps, the deprojected visibility profile\,(c) and the spatially integrated spectrum\,(d). Similar to the continuum emission, the CO $J$= 3-2 momentum zero map shows a double-lobed morphology, which indicates an inner cavity in the CO emission, but the peak-to-peak separation in CO (a preliminary indication of the inner ring radius) is noticeably smaller than that of the continuum emission. The CO emission is also more diffuse than the continuum, extending to $\sim$290\,AU in radius.   Fig.~\ref{fig:co}(c) displays the deprojected CO visibility profile (integrated across the line shape). Compared to the continuum, the null in the visibility profile occurs at a longer baseline length and the side lobes are less prominent,  indicating a smoother CO brightness distribution.

The CO $J$=\,3-2 moment zero map shows depressed emission in the inner $\sim$0.\arcsec3,  which could in principal be explained either by a depleted gas surface density or temperature variations. We argue that temperature variation is unlikely to be the case here. Extremely cold gas inside of 30 AU is physically implausible. For hot gas inside this radius the resulting excitation can indeed produce an optically thin $J$= 3-2 transition, but we do not detect any CO rovibrational emission at 4.7\,$\mum$  with Keck NIRSPEC (probing gas temperature $\geqslant$ 300\,K, e.g. \citealt{Salyk09}). This rules out the possibility of a significant high temperature CO reservoir in the inner disk. Since CO and H$_2$ can survive the intense UV radiation in the cavities of transition disks down to a gas surface density of $10^{-4}$\,g cm$^{-2}$ \citep{Bruderer13}, we assume CO coexists with H$_2$.  We thus consider the most likely case for the depressed CO emission is that the gas is largely depleted inside the cavity.

\section{Modeling Analysis}
\label{sec:model}

The goal of this section is to model the continuum and CO emission from the disk around J1604-2130. Because of differences in optical depths, these two tracers probe different regions of the disk.  At the typical densities of circumstellar disks, the millimeter-wave dust emission is typically optically thin and therefore traces the dust density and temperature in the disk mid-plane. In contrast, the CO emission is expected to be optically thick and therefore traces the gas temperature at larger scale heights. For this reason, we fit models to the dust and CO observations in separate steps.  

First, we construct a disk model that reproduces the broadband SED composed of data from the literature and the ALMA 880\,$\mum$ continuum observations. This enables us to constrain the dust mass surface density and, through radiative transfer calculations, the mid-plane temperature.  Starting from the best fit model for the dust continuum emission, we then fit the CO emission to constrain the parameters that mostly affect the gas emission, such as the gas temperature and the component of the gas velocity along the line-of-sight.

\subsection{Dust emission: model description}
\label{sec:dustmodel}

\begin{figure}[htbp]
\begin{center}
\includegraphics[width=3.3in]{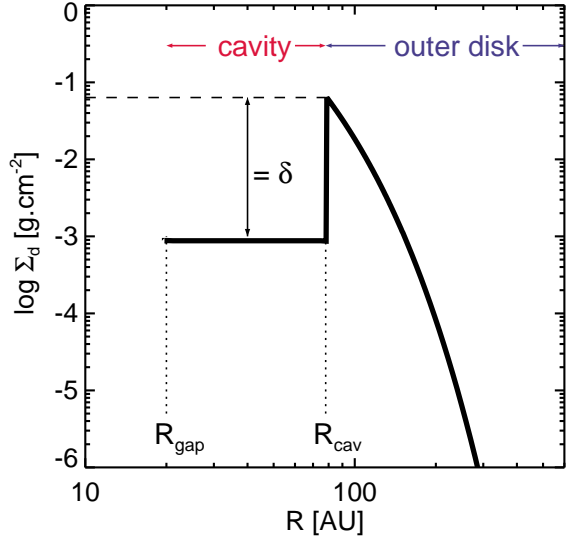}
\caption{Dust surface density model. The disk is divided into three regions: the outer disk (R$>$ R$_{\rm cav}$) follows a similarity solution, while the surface densities inside the cavity ( R$_{\rm gap} <$ R $< $ R$_{\rm cav}$), are scaled down by a factor of $\delta$ compared to the outer disk. The innermost region of the disk ( R $<$ R$_{\rm gap}$)  is empty.  }
\label{fig:sigd}
\end{center}
\end{figure}

We adopt a model for the mass surface density of the dusty disk similar to that 
presented in \citet{Mathews12}, which is briefly summarized here.  In order to 
reproduce the the disk SED and ALMA observations with a minimum 
number of free parameters, we construct a disk model composed of the three radially 
distinct regions (see Figure~\ref{fig:sigd}). 
The outermost disk region, at $R>R_{cav}$, is 
described by the similarity solution of a viscous accretion disk  \citep{Lynden-Bell74} 
adopting the parameterization of \cite{Isella09}, 
\begin{equation}
\label{eq:similarity}
\Sigma(R) =  \Sigma_t \left(\frac{R_t}{R}\right)^{\gamma} \times exp \left\{-\frac{1}{2(2-\gamma)}\left[\left(\frac{R}{R_t}\right)^{2-\gamma}-1\right]\right\},
\end{equation}
where $R_t$ is the radius at which the transition between power law and exponential profiles 
occurs,  $\Sigma_t$ is the surface density at $R_t$, and $\gamma$ defines the slope of the 
surface density profile. At $ R_{gap} < R < R_{cav}$, the disk model has a depleted 
dust region characterized by constant mass surface density $\delta \times \Sigma(R_{cav})$, with 
$\delta <$1.  Finally, to account for the depressed infrared excess for $\lambda <$8\,$\mum$, 
the disk region within $R_{gap}$ is devoid of dust. The dust surface density model is therefore 
defined as
\begin{equation}
\label{eq:sigma_d}
  \Sigma_{dust} (R) =  \left\{
  		\begin{array}{r l} 
                               	0, 																							& 	\mbox{if } R < R_{gap} \\
                               	\delta \times \Sigma (R_{cav}),	&	\mbox{if } R_{gap} \le R < R_{cav} \\
                               	 \Sigma (R),				&	\mbox{if } R > R_{cav} 
		\end{array}
		\right.
\end{equation}

The vertical dust density profile is a Gaussian 
\begin{equation}
\rho(R,Z) = \frac{\Sigma_d(R)}{\sqrt{2\pi}h(R)} e^{-\frac{z^2}{2h^2(R)}}
\end{equation}
in which the scale height $h(R)$ varies with radius as $h(R) = h_0 (R/R_0)^{1+\phi}$.
We set the reference radius $R_0$ to 100\,AU. We choose to parameterize the scale 
height instead of assuming hydrostatic equilibrium because the vertical distribution 
of millimeter grains and gas may differ due to the settling of large dust grains toward the disk mid-plane \citep{Dalessio01}.  

We assume a dust population of astronomical silicates 
and carbonaceous grains, with optical constants as in \cite{Draine03} and 
\cite{Zubko96}, respectively, and at relative abundances as in \cite{Pollack94}. 
Single grain size opacities are calculated using Mie theory 
under the assumption of spherical grains. The single size grain opacities are then integrated 
over the grain size distribution $n(a)$ as described in \cite{Miyake93}. We assume a 
power-law grain size distribution,  $n(a) \propto a^{-3.5}$, from $a_{ \rm min}$= 0.005\,$\mum$ 
to a given $a_{\rm max}$. 

After defining the properties of the dusty disk, we calculate the dust temperature and the 
corresponding emission  using the 2D Monte Carlo radiative transfer code RADMC \citep{Dullemond04}, 
where the stellar radiation field is calculated assuming $T_{\rm eff}$ = 4550\,K, R$_\star$ = 1.41\,R$_\odot$, 
and M$_\star$ = 1 M$_\odot$ \citep{Preibisch02}. 

In summary, the model for the dust continuum emission has 11 free parameters: three that 
describe the gap and cavity of the transitional disk \{$R_{gap}, R_{cav}, \delta$  \}, three that 
define the surface density in the outer disk \{$\Sigma_t$, $R_t$, $\gamma$\},  two that 
characterize the vertical density distribution \{$h_0, \phi$\}, two that define the orientation 
of the disk on the sky, i.e.,  inclination and position angle \{$i$, PA\}, and one ($a_{\rm max}$) that defines
the maximum dust grain size.

\subsection{Dust emission: fitting procedure and results}

Obtaining the best fit model by exploring an 11-D parameter space 
is computationally challenging. Instead,  we reduce the dimensionality of the problem 
by taking into account the fact that the disk SED and the spatially resolved 
880~$\mu$m dust continuum observations  probe two nearly independent 
sub-sets of model parameters. 

The disk infrared emission is optically thick and mainly probes the disk temperature, 
which depends on the vertical profile of the disk through the parameters $h_0$, $\phi$, and 
$R_{gap}$  \citep[e.g.,][]{DAlessio06}. At (sub-)millimeter wavelengths, the disk
emission is optically thin and its spectral slope depends on the dust opacity, which, 
in our model, is controlled by the maximum grain size $a_{max}$ \cite[see, e.g.,][]{Testi14}.
In particular, the optically thin 880\,$\mu$m emission profile probes the surface density of 
millimeter-sized grains, as defined by the parameters 
$\Sigma_t$, $R_t$, $\gamma$, $R_{cav}$,  and $\delta$.
Finally, observations of the molecular line emission are the best tracer of the disk 
orientation. We therefore assume a disk inclination of 6\arcdeg\ and a position angle 
of 77\arcdeg, as derived from the analysis of the CO emission presented in Section~\ref{sec:co_model_result}.
 
To find the disk model that best reproduces the dust
emission observations, we first use the slope of the millimeter SED to constrain the maximum grain size $a_{max}$. 
In the optically thin case, the slope of the dust opacity $\beta$ is related to 
the slope of the flux, $\alpha$, by the relation $\alpha =\beta+2$. 
In the case of J1604-2130, $\alpha =  3.2\pm$0.2, leading to $\beta$ = 1.2 \citep{Mathews12}. 
This corresponds to a maximum grain size of about 0.05\,mm for the dust composition 
and grain size distribution discussed above. The dust opacity at 880\,$\mum$ is 7.26\,cm$^2$ g$^{-1}$. 

We then use the Levenberg-Marquardt $\chi^2$ minimization algorithm (LM) to search 
for the optimum values of $h_0$, $\phi$, and $R_{gap}$ by comparing the 
synthetic disk emission with the measured SED between 1-200\,$\mu$m. 
We note that the IRAC photometric data for J1604-2130 are discrepant from its IRS spectra and the WISE 
photometry results  between 3 and 16\,$\mum$ \citep{Carpenter06, Dahm09, Luhman12}. This discrepancy might suggest variability in the IR disk emission as observed, e.g., in the case of LLRL 31 \citep{Muzerolle09}. Taken at face value, the near-IR excess suggested by the WISE photometry is well fit with an optically thick ring of dust at the temperature of 1500\,K extending from 0.043\,AU to 0.049\,AU, where the former radius is the dust evaporation radius for a star with the properties of J1604-2130 \citep{Isella06}.  Such a warm dust component so close to the star would be distinct from the cool outer disk, whose dust emission peaks at 100\,$\mum$. Thus, fitting the IRAC data only, as was done in \citealt{Mathews12}, should have little effect on other disk parameters except R$_{gap}$.

\begin{figure}[htbp]
\begin{center}
\vspace{-0.5cm}
\includegraphics[width=3.3in]{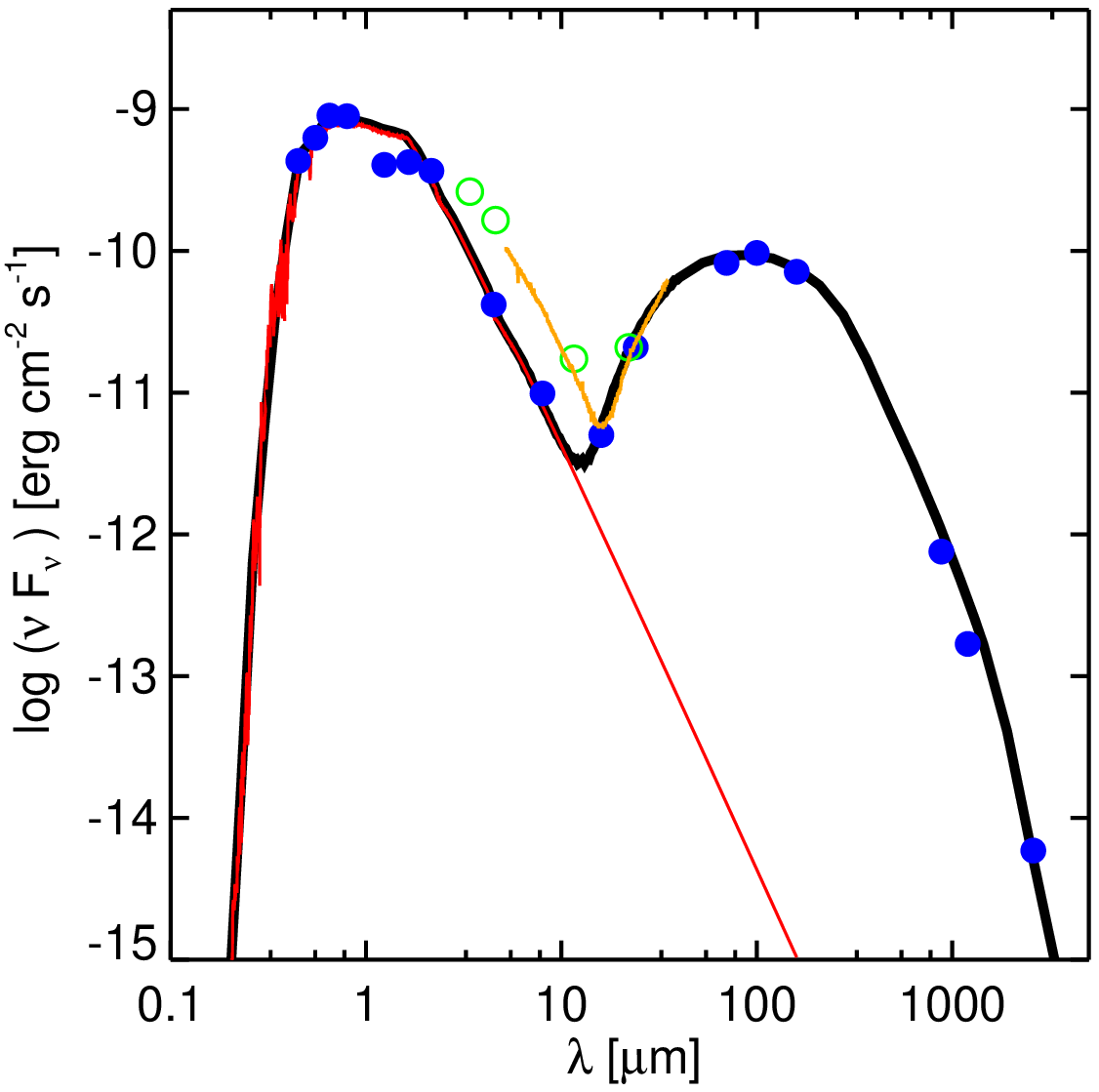}
\caption{Spectral energy distribution and model (see text) for J1604-2130. The red line shows a  Kurucz model atmosphere normalized to the optical and near-IR photometry. The circles are photometric data as follows: the B and R photometry are from USNO-A2.0 \citep{Monet98}, V photometry is from NOMAD \citep{Zacharias05} and I photometry is taken from the DENIS catalogue \citep{DENIS05}.  The BVRI data have been de-reddened with A$_v$ = 0.66 \citep{Carpenter14}. The $2MASS$ $J, H, K$ measurements are from \citet{Cutri03}; $Spitzer$ 4.5,  8 and 16\,$\mum$ photometry is published in \citet{Carpenter06}.  The green open circles are photometry data from WISE catalog. The discrepancy between the WISE and IRAC photometry is discussed in the text.  The IRS spectra (Orange) is taken from \cite{Dahm09}. The MIPS 24\,$\mum$ photometry is from \citet{Carpenter09} while the PACS 70, 100 and 160\,$\mum$ data are from \citet{Mathews13}. The 880\,$\mum$ data point is based on this work. The 1.2\,mm point and 2.6\,mm fluxes are from \citet{Mathews12}.  The solid black line is the best fit model to the star$+$disk system. }
\label{fig:sed}
\end{center}
\end{figure}

In either case and as discussed above, the infrared SED is insensitive to the exact value of the disk 
surface density provided the disk is optically thick to stellar radiation. In this step we therefore assume 
the surface density used by \citet{Mathews12}. We show the best fit model for the SED in Figure~\ref{fig:sed},
for which R$_{gap}$ = 14.6$\pm$2.5\,AU, $\phi$ = 0.6$\pm$0.1,  
and $h_{100 {\rm AU}}$ = 4.0$\pm$0.2\,AU. Fitting the WISE photometry with an optically thick blackbody component results in a larger R$_{gap}$ = 26.5$\pm$2.8\,AU, closer to that derived from CO (discussed in Section 4.4).

As second step, we compare synthetic 880\,$\mu$m images to the ALMA observations to constrain 
$\Sigma_t$, $R_t$, $\gamma$, $R_{cav}$,  and $\delta$. To make a fair comparison with
previous work on transition disks, we adopt $\gamma=1$ \citep{Andrews07, Andrews11}. 

We will discuss the effect that this choice has on the results in \S~\ref{sec:gamma}.

The model fitting  is performed in the Fourier domain adopting the procedure discussed in \cite{Isella09}.
In brief, the synthetic image is Fourier transformed to calculate the synthetic visibilities on 
a regular {\it uv} grid, which are then interpolated at the $uv$ coordinates sampled by the ALMA observations. 
The $\chi^2$, defined as 
\begin{equation}
\label{eq:chi2}
\chi^2 = \sum\limits_{i}[({\rm Re}_{i}^o -{\rm Re}_{i}^m)^2+ ({\rm Im}_{i}^o-{\rm Im}_{i}^m)^2]\cdot w_{i}^o,
\end{equation}
is used as the likelihood estimator. In Equation~\ref{eq:chi2}, Re and Im are the real and 
imaginary part of the complex visibilities; the upper-case ``o" means observation 
and ``m" stands for model. The weight $w$ is calculated as described in \S 2.

We adopt a hybrid method of Levenberg-Marquardt $\chi^2$ minimization algorithm (LM) 
and Markov chain Monte Carlo (MCMC) approach \citep{Pitu07, Isella09}. The LM algorithm is
first used to search for the best fit values, from which the MCMC code is used to obtain 
probability distributions around the LM best fit parameters.  In practice, we launch 150 independent LM
algorithms with random initial values to avoid local minima. Most searches end up with very 
similar values. Then, we carry out 10 independent MCMC calculations starting at the medium value of the 
best fit LM searches. A total number of 5$\times$10$^4$ models were run. 

\begin{figure*}[htbp]
\begin{center}
\vspace{-0.7cm}
\includegraphics[width=7.in]{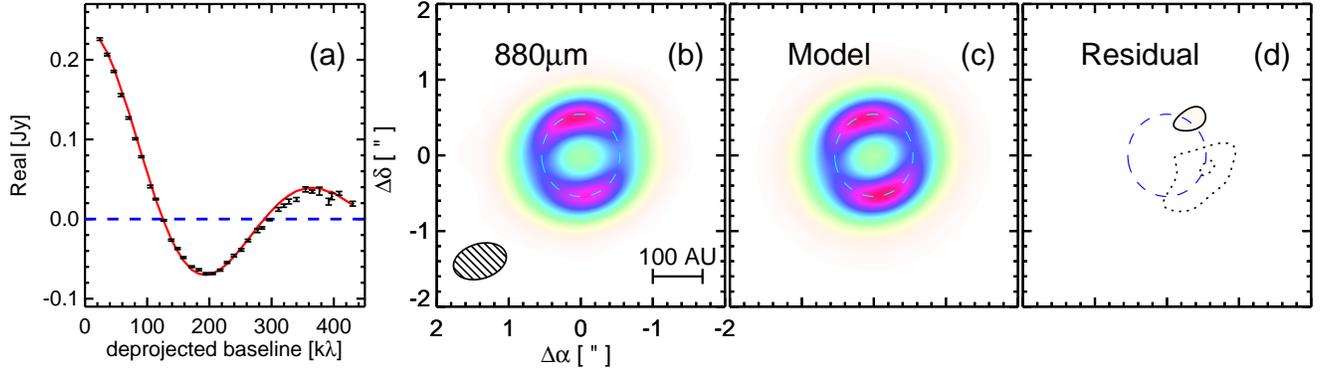}
\caption{ (a) Comparison of the ALMA 880\,$\mum$ continuum visibility profile (black) with the synthetic 
model (red line) summarized in Table~\ref{table:paras}. (b) 880\,$\mum$ continuum map of J1604-2130. 
The cyan dashed circle indicates the truncation radius of dust cavity (R$_{cav}$). (c) 880\,$\mum$ continuum 
model image. (d) 880\,$\mum$ continuum residual map with contours indicating 5 and 
10\,$\sigma$ residuals (solid contours for positive residuals and dotted contours for negative 
residuals).}
\label{fig:cont_model}
\end{center}
\end{figure*}

Table~\ref{table:paras} lists the best fit values for the dust continuum model, which is compared to 
the ALMA 880\,$\mum$ continuum data in Figure~\ref{fig:cont_model}. 
Panel (a) shows the azimuthally averaged visibilities, while panels (b - d) display the
continuum image, model and residual. The residual image is produced by subtracting the model visibilities from those observed, and then imaging the results following the same procedure used for the observations. 
Low level residuals ($\sim$5\,$\sigma$) exist on the west side of the disk and 
suggest an azimuthal variation in the dust density or temperature of {\lapprox}10\%. 

We find that the dust cavity radius is about 79 AU. Within this radius the dust is 
depleted by almost two orders of magnitude. The surface density at the cavity edge 
corresponds to an optical depth of 0.7, and the total mass in dust is 0.16\,M$_{\rm Jup}$.
The transition radius, R$_t$ (see Eq.~\ref{eq:similarity}),  is much smaller than the cavity radius, implying the 
dust surface density at the position of the ring decreases nearly exponentially with 
radius. As a result, the dusty ring is very narrow, with approximately 80\% of the mass
contained between 79\,AU and 114\,AU.

\begin{deluxetable}{llc}
\tabletypesize{\footnotesize}
\tablecolumns{3}
\tablewidth{0pt}
\tablecaption{ Disk parameters derived from 880\,$\mum$ continuum \\
and CO $J=$ 3-2 \label{table:paras}}
\tablehead{
 \multicolumn{3}{c}{Dust disk parameters~~~~~~reduced $\chi^2$ = 2.02}}

\startdata
Cavity radius      &  R$_{\rm cav}$ (AU)       & 78.8$\pm$0.1\vspace{0.05cm}\\
Depletion factor  &  log($\delta$)                          & -1.86$\pm$0.03\vspace{0.1cm}\\
Surface density & $\Sigma$ (R$_{\rm cav})$ (g cm$^{-2}$) &  (9.4$\pm$0.1)$\times10^{-2}$\vspace{0.05cm}\\
Transition radius & R$_t$ (AU)                     &10.6$\pm$0.2\vspace{0.05cm}\\

\hline
\vspace{-0.18cm}\\
 \multicolumn{3}{c}{CO parameters~~~~~~~~~~reduced $\chi^2$ = 1.02}\\
 \vspace{-0.18cm}\\
  \hline
  \vspace{-0.1cm}\\
Cavity radius      &  R$_{\rm CO}$ (AU)       & 31.2$\pm$0.3\vspace{0.05cm}\\
Turbulence factor       & $\xi$                                &  0.25$\pm$0.01\vspace{0.05cm}\\
Temperature at 100AU & T$_0$ (K)                     &45.8$\pm$0.2\vspace{0.05cm}\\
Temperature index  &  $q$                         & -0.64$\pm$0.01\vspace{0.05cm}\\
Position angle     & PA ($^\circ$)                  & 76.8$\pm$0.5
\enddata
\tablecomments{The errors are 1$\sigma$ uncertainty.}
\end{deluxetable}

\subsection{CO emission: model description}

The model used to analyze the ALMA $^{12}$CO $J$=\,3-2 
spectral line observations is similar to that presented in \cite{Isella07}.  
At the typical densities of protoplanetary disks, the rotational transitions of CO are optically 
thick and therefore trace the temperature of the layer characterized by $\tau_{\rm CO} \sim$1
\citep{Beckwith91, Beckwith93, Dutrey96, Isella07}.
Chemical models of disks and direct observations suggest that the optically thick CO 
layer is located far above the mid-plane, where the gas density 
is low and the collisions between dust grains and CO molecules are rare \citep{Woitke09, Walsh12}. 
Thus, a proper calculation of the gas temperature 
requires a fully coupled chemical and line radiative transfer model 
(e.g. \citealt{Kamp04, Woitke09, Walsh12}).  Since this is beyond the scope of this paper, we 
parametrize the CO temperature as a power law, that is, $T_{\rm CO}(R) =  T_{0} \times(R/100 {\rm AU})^{q}$.
Although this choice is arbitrary, it enables us to describe the CO 
temperature independently from that of the dust with a minimum number of free parameters. 

The low-$J$ CO lines generally provide poor constraints 
on the CO column density, and thus on the overall gas content. We therefore fix the gas-to-dust
mass ratio and assume the gas and dust are well mixed. More precisely, the gas density 
is obtained by multiplying the dust density of the best fit model for the continuum emission by 
a constant gas-to-dust mass ratio of 100. The CO number density at each grid point of the disk is then
 calculated by assuming a constant CO/H$_2$ abundance ratio of 5$\times10^{-5}$ \citep{Aikawa96}. 
The critical density of  the CO 3-2 transition is $\sim$10$^4\,$cm$^{-3}$  
from 5-2000\,K (Yang et al. 2010), a value reached in our disk model only in
regions at least five scale heights above the mid-plane.  We can thus safely assume the CO is in Local Thermodynamic Equilibrium (LTE).

We allow for distinct gas and
dust truncation radii as the CO moment zero map reveals 
that the gas disk extends inside the cavity observed via the dust. 
We set an outer edge of the CO disk as that radius where the CO 
temperature reaches 17\,K -- the condensation temperature of CO onto dust grains \citep{Aikawa96}.

For each line-of-sight in the disk, the observed CO line width has contributions from the 
different rotational velocities and thermal plus turbulent broadening. 
The thermal broadening is calculated as $V_{\rm thermal}$=$\sqrt{2k_bT_{\rm CO}/m_{\rm CO}}$, where $k_b$ is the 
Boltzmann constant and $m_{\rm CO}=28$\,m$_{\rm H}$ is the CO molecular mass. The turbulent 
velocity is assumed to be equal to a fraction of the local sound speed, that is, $V_{turb}=\xi c_s$, 
where $\xi$ is a free parameter. Finally, the gaseous disk is assumed to rotate at 
the Keplerian velocity corresponding to a stellar mass of 1~M$_\odot$, and the component along 
the line-of-sight is calculated by assuming a disk inclination angle of 6\arcdeg\ as measured by 
\cite{Mathews12}. 

In summary, the model for the CO emission has five free parameters: 
the CO temperature at 100~AU \{$T_{0}$\}, the radial slope of the CO temperature \{$q$\}, 
the turbulence velocity parameter \{$\xi$\}, the inner radius of the CO disk \{R$_{\rm CO}$\}, 
and the position angle of the disk \{PA\}.

\subsection{{\rm CO} emission: fitting procedure and results\label{sec:co_model_result}}

For any set of free parameters \{$T_{0}$, $q$, $\xi$, R$_{\rm CO}$, PA\} we calculate the 
synthetic CO $J$= 3-2 line emission using the 
ray-tracing code RADLite \citep{Pontoppidan09} under LTE assumption. 
The model cube is convolved to the velocity resolution of the observations and 
re-sampled at the velocity grid of the channel maps. As with the dust continuum emission model, 
the synthetic CO channel maps are Fourier transformed to calculate the theoretical visibilities 
at the {\it uv} coordinates sampled by the data. A minimization on 
the $\chi^2$ surface (Equation~\ref{eq:chi2}) is then performed to find the best fit model. 

Following the dust modeling, we start with 150 independent LM 
runs and random initial values in the search.  However, the 
computational time required to generate synthetic CO data cubes precludes an
efficient sampling of the four dimensional $\chi^2$ surface using a MCMC procedure. 
Instead, we use the Hessian matrix to derive the the uncertainties on the model free parameters 
following the method presented in \citet{Pitu07}.

\begin{figure*}[htbp]
\begin{center}
\vspace{-0.7cm}
\includegraphics[width=7in]{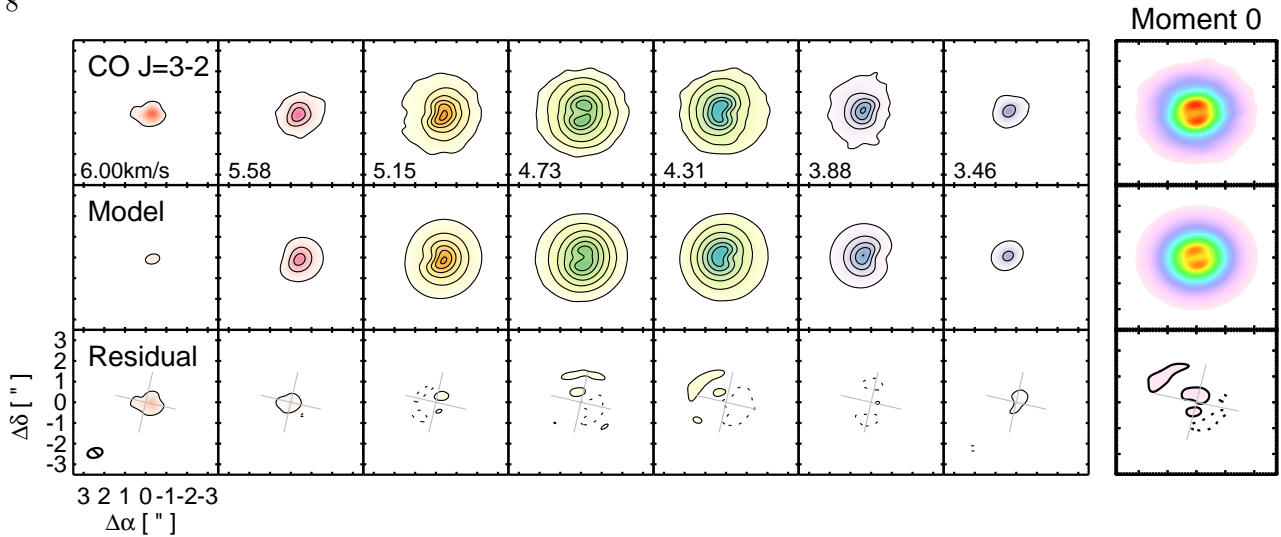}
\caption{Comparison of the observed CO $J$ =3-2 channel and moment zero maps with 
synthetic data from the CO model. The solid contours start at 3$\sigma$ 
(45~mJy beam$^{-1}$) and increase at 10$\sigma$ intervals. Dashed contours indicate 
the -3$\sigma$ level. The color scale of the moment zero maps is the same as that in Figure~\ref{fig:co} (a). }
\label{fig:co_model}
\end{center}
\end{figure*}

The best fit values for the model parameters are listed in Table~\ref{table:paras}, while the comparison
between the synthetic and observed channel maps is presented in Figure~\ref{fig:co_model}.
We find that the CO disk is truncated at an inner radius of 31\,AU, larger than
the gap radius (R$_{\rm gap}$)  inferred for the dusty disk from the SED modeling but half that of the 
partially dust depleted cavity (R$_{\rm cav}$) derived from the spatially resolved 880\,$\mu$m continuum 
observations. Our analysis suggests that the temperature of the CO layer decreases with 
radius as $r^{-0.64}$ starting from a temperature of $\sim$100 K at 31\,AU. The 
CO emission becomes optically thin around 350\,AU, which sets 
the outer radius of the observable CO disk.  Between 80 to 200\,AU the temperature of the 
CO layer is 20-30\,K higher than the mid-plane dust temperature,
consistent with our initial assumption that 
the CO $J$= 3-2 emission mainly arises from a warm molecular layer above the mid-plane. 
 
The residual maps presented in Figure~\ref{fig:co_model} reveal that our model 
reproduces the main kinematic features of the CO emission. Thus, the observations 
are consistent with Keplerian rotation. We derive a turbulent velocity equal to $\sim$25\% of the 
local sound speed, which is within the range predicted by MHD turbulence models \citep{Simon11}. 
Finally, we measure a disk position angle of 77\arcdeg, a value consistent within 
errors with that derived from SMA 880\,$\mum$ continuum and
1.6\,$\mum$ polarized intensity images \citep{Mathews12, Mayama12}.

\subsection{Uncertainties from the choice of $\gamma$\label{sec:gamma}}

\begin{figure*}[htbp]
\begin{center}
\vspace{-0.7cm}
\includegraphics[width=6in]{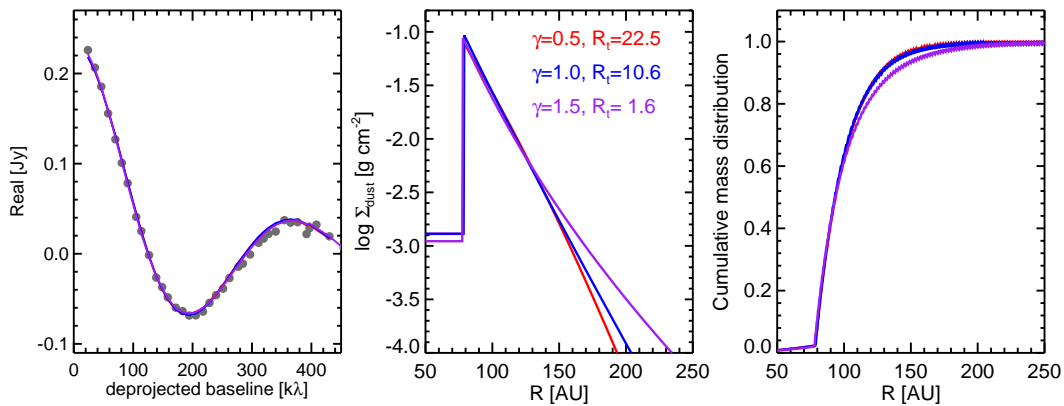}
\caption{Left: Comparison of models with $\gamma$ = 0.5, 1.0 and 1.5 
(Table.~\ref{tab:gamma}) to the 880\,$\mum$ continuum visibility profile (filled grey circles).
All three models clearly produce similar fit to the data. Middle and Right: Surface mass density
and normalized cumulative mass distributions of the models with the three different $\gamma$ assumed. }
\label{fig:gamma}
\end{center}
\end{figure*}

The results obtained so far assume that the disk surface density has 
a radial profile characterized by $\gamma=1$. We relax this assumption 
here to assess its impact on the resulting disk structure. 
To this end, we repeated the dust continuum modeling assuming
$\gamma=0.5$ and $\gamma=1.5$. The best fit models for $\gamma=$0.5, 1, and 1.5
are listed in Table~\ref{tab:gamma}, while Figure~\ref{fig:gamma} shows the comparison 
with data along with the corresponding surface density profiles and cumulative 
mass distributions. 

All three models provide similar quality fits to the 880\,$\mum$ visibilities. 
As expected, we find that $\gamma$ and $R_t$ are degenerate.
Varying $\gamma$ also contributes additional uncertainty to the cavity size 
and depletion factor estimates. Even so, the cavity size remains well constrained under 
the assumption of a sharp truncation at the cavity edge, with an uncertainty of $\sim$1\,AU.  
The depletion factor varies from 0.013 to 0.018.

More generally,  the parameters vary
with $\gamma$ by more than the formal uncertainties derived by the MCMC probability 
distribution sampling discussed above.  Thus the uncertainties in Table~\ref{table:paras} should be used with caution. 
Nevertheless,  the cumulative mass distributions of the three best fit models suggest 
that 80\% of the dust mass traced by the ALMA data is concentrated 
in an annulus extending from $\sim$ 80\,AU to\,120 AU, a result that is 
independent of the exact disk mass surface density profile.

\begin{deluxetable}{cccccc}
\tabletypesize{\footnotesize}
\tablecolumns{5}
\tablecaption{ Continuum model parameters for different $\gamma$ \label{tab:gamma}}
\tablehead{
\colhead{$\gamma$} & \colhead{R$_{cav}$}                  & \colhead{log$\delta$}           &  \colhead{$\Sigma(R_{cav})$} &R$_t$ &$\chi^2_r$\\
                                      &  \colhead{ [ AU ]}                          &                                                   & \colhead{[ g cm$^{-2}$ ] }        & \colhead{[ AU ]}   &}
\startdata
0.5                                &    78.2                             & -1.79                     & $7.8\times10^{-2}$    &    22.5       & 2.029\\
1.0                                &    78.8                             & -1.86                     & $9.4\times10^{-2}$    &    10.6        & 2.024\\
1.5                                &    78.0                           & -1.90                    & $8.8\times10^{-2}$    &    1.6          & 2.025
\enddata
\end{deluxetable}

\section{Discussion}
\label{sec:dis}

\subsection{Comparing J1604-2130 with other transition disks}

The ALMA observations reveal that J1604-2130 disk has a large dust cavity with most of its dust mass concentrated in an narrow annulus ($\sim$35\,AU wide) beyond the dust truncation radius. 
Here we compare the mass distribution of J1604-2130 with other transition disks. 

We define a characteristic width, $\Delta W$, by 
\begin{equation}
\label{eq:width}
\int_{R_{cav}}^{R_{cav}+\Delta W} \Sigma(r) \times2\pi r dr = 0.8 \times\int_{R_{cav}}^{+\infty} \Sigma(r) \times2\pi r dr
\end{equation}

The average radius of the annulus holding 80\% of disk mass is then $\overline{R}$=$R_{cav}+\Delta W/2$. The $\Delta W$/$\overline{R}$ ratio can be used as an tracer 
of the compactness of the mass distribution. In Figure~\ref{fig:cummass}(b), we compare $\Delta W$/$\overline{R}$ for J1604-2130 with that for the transition disks studied in \citet{Andrews11}.  The surface density in all 13 disks was determined by analyzing resolved  880\,$\mum$ continuum visibilities using similarity solution with $\gamma$ fixed at 1. Thus, all of the results are expected to represent the mass distributions of dust grains over similar sizes. Interestingly, J1604-2130 has the largest dust cavity among the thirteen disks but is also the most compact as measured via $\Delta W/\overline{R}$.  It is worth to mention that some of  the transition disks might actually be more compact than they appeared since their SMA observations could be affected by decorrelation that will spread the dust emission. 

This high concentration of dust and the large cavity radius is difficult to understand from viscous evolution alone because an expanding disk should produce a more diffuse mass distribution, unless a very large cavity can be opened early in the life time of the disk, $\sim$10$^5$\,yr \citep{Isella09}.  On the other hand, transition disks are clearly going through significant changes, so mechanisms other
than viscosity are likely to be important in shaping the mass surface density at this particular evolutionary stage. We discuss possible scenarios for the evolution of the J1604-2130 transition disk next.

\subsection{Formation of the J1604-2130 transition disk}

The main result from the model fits to the dust continuum and CO emission is that the 
radius of the gas cavity (31\,AU) is roughly half that of the dust (79\,AU). Similar results are found in the transition disk IRS 48, 
where CO shows a 20\,AU cavity while the dust is truncated at $\sim$45\,AU \citep{Bruderer14}.

Such differences in the gas and dust distributions have implications for the formation mechanisms of transitions disks. It is known that gas exists inside the dust cavity in some transition disks.  \citet{Salyk09} detected 
rovibrational CO lines near 4.7\,$\mum$ in nine out of fourteen transitional disks,  suggesting that high temperature gas often exists inside the cavity. Several transition disks show rotational CO emission inside dust cavity, indicating that cool gas is present (e.g. \citealt{Pietu06, Isella10, Tang12, Casassus13,Perez14}). However, these observations were unable to determine if there is a cavity/gap in gas. Only in the case of GM Tau was CO inferred to be depleted within 20\,AU based on the lack of high velocity wings in the rotational transitions of CO isotopologues \citep {Dutrey08}.

We discuss here the possible formation mechanisms for the J1604-2130 transition disk, using the separate truncation edges in the dust and gas along with other observational properties as constraints. We consider: (1) grain growth inside the cavity \citep{DAlessio06},  (2) mass-loss in a wind driven by photoevaporation or magneto-rotational instability (MRI) \citep{Alexander06a, Alexander06b, Suzuki09}, and (3) tidal interactions with companions \citep{Bryden99, Crida06}. 

\textit{Grain growth.} For J1604-2130, the clear detection of a large gas cavity indicates that grain growth is unlikely to be the dominant clearing mechanism, because 
even small amounts of gas containing CO are expected to survive inside a cavity where dust grains have grown to large size \citep{Bruderer13}.  Thus, the lack of CO emission inside of $\sim$ 31\,AU cannot be caused by pure grain growth.

\textit{Photoevaporation.}  Another possible mechanism for gas and dust removal is a wind driven by photoevaporation. However, two observational properties of J1604-2130 are inconsistent with the predictions of current photoevaporation models.

The first problem is the large separation between the gas and dust truncation radii. Disk photoevaporation preferentially removes the gaseous disk and sub-micron sized grains entrained in the gas flow.  The inner edge of the gas disk moves steadily outwards and the strong inward pressure gradient at the evaporation radius sweeps up all dust grains smaller than a few millimeters in size. Only very large grains ($\ge$ cm-sized) might be able to survive inside the gas-free regions if not removed by other means.   In this case, the gaseous and dusty disk components are both truncated at the inner rim of the outer disk \citep{Alexander07}.  Thus, for either the grain growth or photoevaporation
scenarios it is difficult to produce the large separation between the characteristic gas and mm-sized dust radii in J1604-2130. 

The second difficulty concerns the high mass surface density of J1604-2130. Photoevaporation winds continuously remove material from the disk surface. When a gap as large as tens of AU is opened, the surface density of the outer disk is expected to be less than 0.1\,g cm$^{-2}$ \citep{Alexander06b, Alexander07, Gorti09}. The gas surface density of J1604-2130 at the truncation edge is $\sim$10\,g cm$^{-2}$ (assuming a gas-to-dust ratio of 100), far too massive to be explained by photoevaporation.

Tidal interactions with low mass companions. Here, interactions between brown dwarf companions and the disk are unlikely to be the cause of the J1604-2130 cavity.\citet{Kraus08} have ruled out the presence of a close companion with aperture masking interferometry (0.06\,M$_\odot$ down to $\sim$2\,AU, and 0.01\,M$_\odot$ down to $\sim$9 AU); while \citet{Ireland11} have placed upper limits on companion masses of 0.07-0.005\,M$_\odot$ at separations from 60-300 AU using adaptive optics.  We thus compare the CO and dust radial structure with the predictions of disk tidal interactions with planets. 

The disk-planet interaction scenario can naturally explain the large morphological differences in the gas and dust emission in J1604-2130. Recent disk-planet dynamical models (e.g. \citealt{Pinilla12};  \citealt{Zhu12}) have begun to incorporate dust coagulation and fragmentation processes in hydrodynamical simulations. One of the key features of these new models is that the difference between the planet orbital radius and the peak of surface density of mm-sized particles may be as large as tens of AU if the planet is sufficiently massive. A giant planet not only opens a gap in the disk, but also produces a local pressure maximum where dust particles within some size range can be trapped. The planet orbit and local pressure maximum separation depends on the mass of the planet -- the more massive the planet, the larger the separation. For example, \citet{Pinilla12} showed that the local pressure maximum can be located at more than twice the planet orbital radius for a 9\,M$_{\rm Jup}$ planet. 
 
\begin{figure*}
\begin{center}
\vspace{-0.5cm}
\includegraphics[width=6in]{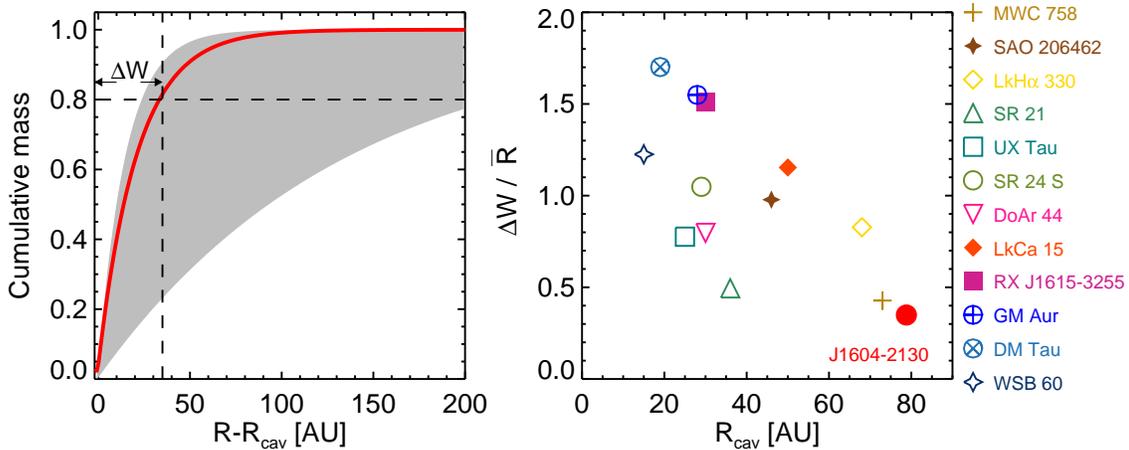}
\caption{Left: The normalized cumulative mass distribution profile of J1604-2130 (red solid line) as compared to the range  of distributions observed in twelve transition disks by \citet{Andrews11} (shaded region). It can be seen that J1604-2130 has a high mass concentration beyond its dust cavity. We also illustrate the characteristic width $\Delta W$ as the width of the annulus including 80\% of the
disk mass. Right: Comparison of dust cavity sizes with the compactness of the dust ring in the thirteen transition disks plotted at left. The compactness of the dust ring is defined as the ratio between the width of the annulus including 80\% of the dust mass and the mean radius of the dust ring, i.e., $\Delta W/\overline{R}$}. J1604-2130 has the largest dust cavity among the thirteen disks, while its mass distribution is among the most compact.
\label{fig:cummass}
\end{center}
\end{figure*}

A planet-induced disk pressure trap is also consistent with the high concentration of mm-sized dust in J1604-2130. Because the efficiency of dust trapping is size-dependent, the general observable outcome would be that the large grains are trapped into the local pressure maximum while small grains and gas should persist in a more diffuse distribution \citep{deJuanOvelar13}. As shown in Figure~\ref{fig:cummass}, J1604-2130 has most of its mm-sized dust mass concentrated in a highly radially confined ring.  Further, \citet{Mayama12} reported that the 1.6\,$\mum$ polarized intensity images (tracing micron-sized particles) that peak at $\sim$63\,AU, which is 15\,AU closer to the central star than the 880\,$\mum$ emission. The 1.6$\mum$ data also show that scattered light extends inwards to $\sim$40\,AU (smaller separations lie within their saturation radius).

\subsection{Pressure trap and evolution}

In the ALMA 880\,$\mum$ continuum and CO $J$=3-2 emission maps, J1604-2130 appears to be regular and largely azimuthally symmetric -- but with the dust highly concentrated in a narrow ring.
A similar morphology has recently been reported  in the evolved circumbinary disk around V4046 Sgr \citep{Rosenfeld13}.
In contrast, large azimuthal asymmetry features have been observed in several transitional disks \citep{Brown09, Regaly12, Isella13, vanderMarel13, Perez14}, possibly caused by extended vortices created by planet-disk tidal forces or large viscosity gradients.  Intriguingly, the two types of dust morphologies (azimuthal asymmetry and symmetry) are both consistent with the idea of a pressure trap, but are expected at different evolutionary stages.

\citet{Fu14}, for example, show that  the emergence and lifetime of the vortices created by planets strongly depend on the disk viscosity -- with vortices typically diffusing into a ring-like structures at later stages --while simulations of vortices caused by large viscosity gradients demonstrate that the vortices and rings appear alternately (see Figure 3 in \citealt{Regaly12}).  To make any further statement about the nature of these dust concentrations in disks, other manifestations of pressure trap must be observed, such as the expected radial variation of the dust-to-gas mass ratio, as well as disks at different evolutionary stages.

\subsection{Dust inside the cavity}

Transitional disks were first identified through the dip in their infrared SED, suggesting that warm dust near the central star is substantially depleted \citep{Skrutskie90}. Since the infrared dust emission is usually optically thick, only lower limits to dust mass can be derived from the SED \citep{Andrews11}. In contrast, the millimeter wavelength dust emission probes much larger column densities, but spatially resolved millimeter images have so far lacked the sensitivity 
to constrain the dust density inside the cavity.  There is one case, LkCa 15, in which a small amount of 870\,$\mum$ continuum emission has been detected inside the dust cavity, leading to an estimated surface density $\sim$5 times lower than the regions immediately beyond the dust cavity \citep{Andrews11b}. Our observation of J1604-2130 also shows detectable emission (at $\sim$10$\sigma$) inside the cavity with a surface density of mm-sized particles nearly $\sim$100 times lower than that of the disk beyond the dust cavity.  In both cases, the depletions estimated from sub-mm images are significantly higher than the 10$^{-6}$-- 10$^{-5}$ typical lower limit from the transition disk SED analysis \citep{Andrews11}. We stress, however, that the two depletion factors are difficult to quantitatively compare because the IR emission mostly arises from small dust grains ($\sim$$\mum$ size) in the disk atmosphere while the sub-mm emission is mainly from large grains ($\sim$mm-sized) close to the mid-plane of the disk.

\vskip -0.2in
\subsection{Outer disk radius}

The CO $J$=3-2 emission of J1604-2130 appears to extend to a radial distance much larger than the associated 880\,$\mum$ continuum (see Figures~\ref{fig:cont} and \ref{fig:co}). Similarly large discrepancies between continuum and gas emission have been observed in several other protoplanetary disks \citep{Pietu05, Isella07, Panic09, Andrews12, Rosenfeld13b}.  It has being debated whether the apparent size discrepancy reflects actual differences in the outer radii of dust and gas in these disks.  \citet{Hughes08} suggested that the apparent discrepancy may be due to optical depth effects, and demonstrated that models with a tapered exponential edge in the surface density profile could better reproduce the dust and gas observations. There are some cases, however, where tapered surface density profiles are unable to simultaneously reproduce the CO and dust observations \citep{Panic09, Andrews12, Rosenfeld13b}. In such cases, a radially varying gas-to-dust mass ratio is required to explain the observations. For the case of J1604-2130, we find that a tapered surface density profile and radially invariant gas-to-dust ratio is consistent with the current CO and continuum observations. Future high spatial resolution observation of optically thin transitions (e.g. CO isotopologue lines) are necessary to distinguish if the outer edge of gas and dust in J1604-2130 are truncated at different radii.

\vskip 0.2in
 \section{Summary}
\label{sec:summary}

In this paper, we have studied the radial structure of  the transitional disk J1604-2130 with sensitive ALMA data of the 880\,$\mum$ continuum and CO $J$ = 3-2 line emission. The key conclusion of our analysis are:\begin{itemize}
  \item Both the dust continuum and CO gas show central cavities in the  synthesized images but the gas truncation radius is about half that of the dust (R$_{\rm CO}$ = 31\,AU vs. R$_{\rm cav}$ = 79\,AU).  The presence of a large gas cavity rules out the possibility of dust growth as the main mechanism for the central depletion. A large separation in the edges of gas and dust is not predicted by photoevaporation models, but is expected for massive planet-disk interactions.
   \item Dust inside the cavity is detected in the 880\,$\mum$ continuum, but with a surface density 100 times lower than that just beyond the dust truncation radius. The estimated dust mass inside the dust cavity is  $\sim$1.1M$_{\bigoplus}$.
   
  \item 80\% of the dust mass is concentrated in an annulus extending between 79 and 114\,AU in radius.  This  morphology is qualitatively consistent with the accumulation of dust grains in a local pressure bump such as that generated by the dynamical interaction between a gas giant planet and the disk.

\end{itemize}

\bigskip

\begin{acknowledgments}
We thank Crystal Brogan and Steve Myers for their assistance with the data
reduction. The National Radio Astronomy Observatory is a facility of the
National Science Foundation operated under cooperative agreement by Associated
Universities, Inc. This paper makes use of the following ALMA data:
ADS/JAO.ALMA\#2011.0.00526.S. ALMA is a partnership of ESO (representing its
member states), NSF (USA) and NINS (Japan), together with NRC (Canada) and NSC
and ASIAA (Taiwan), in cooperation with the Republic of Chile. The Joint ALMA
Observatory is operated by ESO, AUI/NRAO and NAOJ. A.I. and J.M.C. acknowledge
support from NSF awards AST-1109334 and AST-1140063. K.Z. and G.A.B. gratefully
acknowledge funding provided by NSF award AST-1109857 and NASA grant NNX11AK86G.
\end{acknowledgments}


\bibliographystyle{apj}
\bibliography{ms}

\end{document}